\documentstyle[11pt,twoside]{article}
\begin{document}\hbadness=10000\thispagestyle{empty}
\pagestyle{myheadings}
\markboth{H.-Th. Elze}
{Fluid Dynamics of Relativistic Quantum Dust}
\title{{\bf Fluid Dynamics of Relativistic Quantum Dust}}
\author{$\ $\\
{\bf Hans-Thomas Elze}\\ $\ $\\
Universidade Federal do Rio de Janeiro, Instituto de F\'{\i}sica\\
\,Caixa Postal 68.528, 21945-970 Rio de Janeiro, RJ, Brazil }
\vskip 0.5cm
\date{December 2001}
\maketitle
\vspace{-8.5cm}
\vspace*{8.0cm}
\begin{abstract}{\noindent
The microscopic transport equations for free fields are solved
using the Schwinger function. Thus, for general initial conditions,
the evolution of the energy-momentum tensor is obtained, incorporating
the quantum effects exactly. The result for relativistic fermions differs
from classical hydrodynamics, which is illustrated for Landau and Bjorken 
type initial conditions in this model of exploding primordial matter. 
Free fermions behave like classical dust concerning hydrodynamic 
observables. However, quantum effects which are present 
in the initial state are preserved.  
\\
\noindent
PACS numbers: 
}\end{abstract}
Often the complicated time dependent dynamics of quantum many-body systems or fields is
approximated by a perfect fluid model. Since the seminal work by Fermi and Landau this
approach has been applied successfully, in order to study global features, such as multiplicity distributions
and apparently thermal transverse momentum spectra of produced particles, in high-energy collisions of strongly
interacting matter \cite{Fermi,Landau,Bj,QMs}. Similarly, the hydrodynamic approximation is often invoked in
astrophysical applications and cosmological studies of the early universe \cite{Weinberg}

Recently it has been shown that a free scalar field indeed behaves like a perfect fluid in
the semiclassical (WKB) regime \cite{DomLev00}. More generally, the mechanisms of quantum
decoherence and thermalization in such systems which can be described hydrodynamically, i.e.
the emergence of classical deterministic evolution from an underlying quantum field theory,
are of fundamental interest \cite{WHZ90,I95,Lisewski99,BrunHartle99}.

The limitations of the fluid picture, however, have rarely been explored in the
microscopic or high energy density domain. Difficulties reside in the derivation
of consistent transport equations and in the amount of computation required
to find realistic solutions; see Refs.\,\cite{EH}, for example, for a review
and recent progress concerning selfinteracting scalar particles and the quark-gluon
plasma, respectively. More understanding of related hydrodynamic behavior, if any,
seems highly desirable.

Presently, we study the relation between relativistic hydrodynamics
and the full quantum evolution of a free matter field. In the absence of
interactions, decoherence or thermalization may be present in the initial
state, corresponding to an impure density matrix, but is followed by unitary
evolution. We consider this as a ``quantum dust'' model of the expansion of matter
originating from a high energy density preparation phase, which the Landau and Bjorken
models describe classically \cite{Landau,Bj}.

Our approach is independent of the nature of the field, as long as it obeys a
standard wave equation. To be definite, we choose to work with
Dirac fermions and comment about neutrinos later.
We introduce the spinor Wigner function, i.e., a (4x4)-matrix depending on
space-time and four-momentum coordinates:
\begin{equation}\label{Wigner}
W_{\alpha\beta}(x;p)\equiv\int\frac{\mbox{d}^4y}{(2\pi )^4}
e^{-ip\cdot y}
\langle :\bar\psi_\beta (x+y/2)\psi_\alpha (x-y/2):\rangle
\;\;, \end{equation}
where the expectation value refers to the (mixed) state of the system;
without interactions, the vacuum plays only a passive role and, therefore,
is eliminated by normal-ordering the field operators.

All observables can be expressed in terms of the Wigner function here.
In particular, the (unsymmetrized) energy-momentum tensor:
\begin{equation}\label{Tmunu}
\langle :T_{\mu\nu}(x):\rangle\equiv i\langle :\bar\psi (x)\gamma_\mu
\stackrel{\leftrightarrow}{\partial}_\nu\psi (x):\rangle
=\mbox{tr}\;\gamma_\mu\int\mbox{d}^4p\;p_\nu W(x;p)
\;\;, \end{equation}
where $\stackrel{\leftrightarrow}{\partial}\equiv\frac{1}{2}(\overrightarrow\partial -\overleftarrow\partial )$
and with a trace over spinor indices (conventions as in \cite{VGE87}). Furthermore,
the dynamics of $W$ reduces to the usual phase space description in the classical limit \cite{VGE87}.

Propagation of the free fields entering in Eq.\,(\ref{Wigner}) from one
time-like hypersurface to another is described by the Schwinger function.
It is the solution of the homogeneous Dirac equation,
$[i\gamma\cdot\partial_x-m]S(x,x')=0$, for the initial condition
$S(\vec x,\vec x',x^0=x'^0)=-i\gamma^0\delta^3(\vec x-\vec x')$. Thus,
$\psi (x)=i\int\mbox{d}^3x'S(x,x')\gamma^0\psi (x')$, and similarly for the adjoint.
An explicit form is:
\begin{equation}\label{Schwinger}
iS(x,x')=iS(x-x'\equiv\Delta)=(i\gamma\cdot\partial_{\Delta}+m)
\int\frac{\mbox{d}^3k}{(2\pi )^32\omega_k}\left (
e^{-ik_+\cdot\Delta}-e^{-ik_-\cdot\Delta}\right )
\;\;, \end{equation}
where $k_\pm\equiv (\pm\omega_k,\vec k)$ and $\omega_k\equiv (\vec k^2+m^2)^{1/2}$.

Making use of Eqs.\,(\ref{Wigner}) and (\ref{Schwinger}),
we relate the Wigner function at different times, $t=x^0,x'^0$:
\begin{equation}\label{Wevolution}
W(x;p)=\int\frac{\mbox{d}^4k}{(2\pi )^3}e^{-ik\cdot x}
\delta^\pm (p_k^+)\delta^\pm (p_k^-)
\int\mbox{d}^3x'e^{ik\cdot x'}\int\mbox{d}p'^0
\Lambda(p_k^+)\gamma^0W(x';p')\gamma^0\Lambda(p_k^-)
\;\;, \end{equation}
where $p_k^\pm\equiv p\pm\frac{k}{2}$, $\delta^\pm (q)\equiv\pm\delta(q^2-m^2)$ (for $q^0/|q^0|=\pm 1$),
$\Lambda (q)\equiv\gamma\cdot q+m$, and $p'^\mu\equiv (p'^0,\vec p)$.

The Eq.\,(\ref{Wevolution}) implies that the Wigner function obeys a generalized mass-shell
constraint and a proper free-streaming transport equation:
\begin{eqnarray}\label{constraint}
[p^2-m^2-\frac{\hbar^2}{4}\partial_x^2]\;W(x;p)&=&0 \;\;, \\ [2ex]
\label{transport} p\cdot\partial_x\;W(x;p)&=&0 \;\;,
\end{eqnarray} separately for each matrix element. The reinserted
$\hbar$ indicates the important quantum term in the equations,
which otherwise have the familiar classical appearance.

Thus Eq.\,(\ref{Wevolution}) presents an integral solution of the microscopic transport equations
for a given initial Wigner function. Furthermore, a semiclassical approximation of the Schwinger function
may be used to generate an integral solution of the corresponding classical transport problem.

Next, we decompose the Wigner function with respect to the standard basis of
the Clifford algebra, $W={\cal F}+i\gamma^5{\cal P}+\gamma^\mu {\cal V}_\mu
+\gamma^\mu\gamma^5{\cal A}_\mu +\frac{1}{2}\sigma^{\mu\nu}{\cal S}_{\mu\nu}$,
i.e., in terms of scalar, pseudoscalar, vector, axial vector, and
antisymmetric tensor components. The functions,
${\cal F}\equiv\frac{1}{4}\mbox{tr}\;W$\,,
${\cal P}\equiv-\frac{1}{4}i\;\mbox{tr}\;\gamma^5 W$\,,
${\cal V}_\mu\equiv\frac{1}{4}\mbox{tr}\;\gamma_\mu W$\,,
${\cal A}_\mu\equiv\frac{1}{4}\mbox{tr}\;\gamma_5\gamma_\mu W$\,,
and ${\cal S}_{\mu\nu}\equiv\frac{1}{4}\mbox{tr}\;\sigma_{\mu\nu}W$\,,
which represent physical current densities,
are real, due to $W^\dagger =\gamma^0W\gamma^0$ \cite{VGE87}. They individually obey
Eqs.\,(\ref{constraint}) and (\ref{transport}).

We assume ${\cal P}=0={\cal A}_\mu$, i.e., we consider a spin saturated system for simplicity
\cite{neutrino}. Then, using the `transport equation' which follows directly from the
Dirac equation applied to $W$, $[\gamma\cdot (p+\frac{i}{2}\partial_x)-m]W(x;p)=0$, and
decomposing it accordingly, the following additional relations among the remaining densities
are obtained:
\begin{eqnarray}\label{vector}
{\cal V}^\mu (x;p)&=&\frac{mp^\mu}{p^2}{\cal F}(x;p)
\;\;, \\ [2ex]
\label{tensor}
{\cal S}^{\mu\nu}(x;p)&=&\frac{1}{2p^2}(p^\nu\partial_x^\mu -p^\mu\partial_x^\nu ){\cal F}(x;p)
\;\;. \end{eqnarray}
Note that ${\cal S}^{\mu\nu}$ is intrinsically by one order in $\hbar$ smaller than the
other two densities.

We conclude that presently the dynamics of the system is represented completely by the scalar phase space
density ${\cal F}$. Using Eqs.\,(\ref{Tmunu}) and (\ref{vector}), we obtain in particular:
\begin{equation}\label{TF}
\langle :T^{\mu\nu}(x):\rangle =4m\int\mbox{d}^4p\;\frac{p^\mu p^\nu}{p^2}{\cal F}(x;p)
\;\;, \end{equation}
which is symmetric and conserved, $\partial_\mu T^{\mu\nu}(x)=0$, on account of Eq.\,(\ref{transport}).
Furthermore, this implies the `equation of state':  
\begin{equation}\label{eos}
\langle :T^{00}(x):\rangle -\sum_{i=1}^3\langle :T^{ii}:\rangle 
=4m\int\mbox{d}^4p\;{\cal F}(x;p)
\;\;, \end{equation}
which relates energy density and pressure(s). However, applying Eq.\,(\ref{constraint}), we find that this relationship evolves in a wavelike manner, driven by off-shell contributions to the evolving ${\cal F}$:
\begin{equation}\label{eoswave}
\partial_x^2\langle :T_\mu^\mu (x):\rangle =16m\int\mbox{d}^4p\;
(p^2-m^2){\cal F}(x;p)
\;\;. \end{equation}
This differs from classical hydrodynamics with a fixed functional form of the equation of state.  
Eqs.\,(\ref{TF})-(\ref{eoswave}) hold independently of the initial state, of course, if  
it evolves without further interaction.  

Making use of  Eq.\,(\ref{Wevolution}) in Eq.\,(\ref{TF}), we now calculate the
energy-momentum tensor at any time in terms of the initial scalar density. Employing the decomposition
of the Wigner function and commutation and trace relations for the $\gamma$ matrices, as well as
Eqs.\,(\ref{constraint})-(\ref{tensor}), we obtain:
\begin{eqnarray}
&\;&\langle :T^{\mu\nu}(x):\rangle =8m\int\dots\int
\frac{p^\mu p^\nu}{p^2}
\left (p^2+\frac{m^2}{p'^2}\left (p^0p'^0+\vec p^{\;2}\right )
-\frac{1}{4p'^2}\left ((p^0k^0)^2-\vec p^{\;2}\vec k^2 \right .
\right . \nonumber
\\[1ex]\label{TmunuEvolution}
&\;&\left . \;\;\;\;\;\;\;\;\;\;\;\;\;\;\;\;\;\;\;\;\;\;\;\;\;\;\;\;
\;\;\;\;\;\;\;\;\;\;\;\;\;\;\;\;\;\;\;\;\;\;\;\;
+p^0p'^0k^2+\frac{p^0}{p'^0}\left .(k^0)^2p^2\right )
\right ){\cal F}(x';\vec p,p'^0)
\;\;, \end{eqnarray}
where $\int\dots\int\equiv (2\pi )^{-3}
\int\mbox{d}^4p\int\mbox{d}^4k\;e^{-ik\cdot x}
\delta^\pm (p_k^+)\delta^\pm (p_k^-)
\int\mbox{d}^3x'e^{ik\cdot x'}\int\mbox{d}p'^0$; we also made use of partial integrations and
the $\delta$-function constraints. The three terms on the right-hand side stem from
the scalar, vector, and antisymmetric tensor
components of the initial Wigner function, respectively. 
 
If the initial distribution is an isotropic function of the three-momentum,
then $T^{\mu\nu}$ is diagonal at all times, implying that the absence of flow 
in the initial state will be preserved.

Indeed, we expect the (non-)flow features of the initial
distribution to be preserved during the evolution, due to the
absence of interactions. Kinetic energy from microscopic
particle degrees of freedom will not be converted into collective
motion. An interesting question is, how the classical hydrodynamic
acceleration of fluid cells due to pressure gradients arises in
our present model after coarse graining
\cite{WHZ90,I95,Lisewski99,BrunHartle99}. We do not pursue this at
present. Recalling earlier work on the hydrodynamic 
representation of quantum mechanics, e.g. Refs.\,\cite{Madelung},
and recently deduced classical fluid behavior of quantum fields in
WKB approximation \cite{DomLev00}, we study the full 
quantum effects here.

We consider the exact evolution of $T^{\mu\nu}$,   
assuming a particle-antiparticle symmetric initial state. This is believed to hold,
for example, close to midrapidity in the center-of-mass frame of central high-energy collisions \cite{Bj,QMs}.
It implies that the initial ${\cal F}$ is an even function of
the energy variable, ${\cal F}(x';\vec p,p'^0)={\cal F}(x';\vec p,-p'^0)$.  
While Eq.\,(\ref{TmunuEvolution}) allows general initial conditions, we follow 
the implicit on-shell assumption in classical hydrodynamic models:
\begin{equation}\label{onshell}
{\cal F}(x';\vec p,p'^0)=(2\pi )^{-3}m\;\delta (p'^2-m^2)\left (
\Theta (p'^0)F(x';\vec p,p'^0)+\Theta (-p'^0)F(x';\vec p,-p'^0)\right )
\;\;. \end{equation}
Fermion blackbody radiation is described by
$F(x';\vec p,p'^0)\equiv f(p'^0/T(x'))$,
where $T$ denotes the local temperature, and with $f(s)\equiv (e^s+1)^{-1}$; this is easily
illustrated with the help of Eqs.\,(\ref{TF}) and (\ref{onshell}). 

Implementing Eq.\,(\ref{onshell}), we obtain the simpler result:
\begin{eqnarray}\label{Tmunuonshell}
&\;&\langle :T^{\mu\nu}(x):\rangle =\int\frac{\mbox{d}^3x'\mbox{d}^3p}{(2\pi )^3}\frac{\mbox{d}^3k}{(2\pi )^3}
\;\frac{p^\mu p^\nu\cos [\vec k\cdot (\vec x-\vec x')]}{\omega_p\omega_+\omega_-}
F(x';\vec p,\omega_p)
\\ [1ex]
&\;&\cdot\{ ((\omega_++\omega_-)^2-\vec k^2 )\cos [(\omega_+-\omega_-)t]
-((\omega_+-\omega_-)^2-\vec k^2)\cos [(\omega_++\omega_-)t] \}
\;\;, \nonumber \end{eqnarray}
where $t\equiv x^0-x'^0$, $\omega_p\equiv (\vec p^2+m^2)^{1/2}$, $\omega_\pm\equiv ((\vec p\pm\vec k/2)^2+m^2)^{1/2}$;  
furthermore, $p^0\equiv\frac{1}{2}|\omega_+\pm\omega_-|$, with ``+'' when multiplying the first and 
``$-$'' when multiplying the second term of the difference, respectively. 
Depending on geometry and initial state, further
integrations can be done analytically.

Consider a $(1+1)$-dimensional system for illustration, 
assuming that the particles are approximately massless, i.e. $\omega_p\approx |p|$, and that 
$F$ is even in $p$ (no flow).   
Specializing to a Landau type initial condition, the distribution is 
prepared on a fixed timelike hypersurface at $t=0$ \cite{Landau}. We    
find the ultrarelativistic equation of state for the only 
nonvanishing components of $T^{\mu\nu}$, $\epsilon\equiv T^{00}=T^{11}\equiv P$ ($d=1+1$), 
which are calculated as a momentum integral following Eq.\,(\ref{Tmunuonshell}): 
\begin{eqnarray}\label{Landau} 
T^{00}(x,t)&=&2\int\frac{\mbox{d}p}{2\pi}|p|{\displaystyle\Bigl (
F(x-t;|p|)+F(x+t;|p|)\Bigr )}
\\ [1ex] \label{LandauWave} 
&=&\frac{1}{2}\left (T^{00}(x-t,t=0)+T^{00}(x+t,t=0)\right ) 
\;\;, \end{eqnarray} 
i.e., a superposition of wavelike propagating momentum contributions 
in accordance with Eq.\,(\ref{eoswave}). 

Similarly, a Bjorken type initial condition can be specified on a surface of 
constant proper time \cite{Bj}. A transformation of Eq.\,(\ref{Landau}) 
to space-time rapidity and proper time coordinates yields:  
\begin{equation}\label{Bjorken} 
T^{00}(y,\tau )=2\int\frac{\mbox{d}p}{2\pi}|p|{\displaystyle\Bigl (
F(-\tau_0 e^{-y/2+\ln \tau /\tau_0};|p|)+F(\tau_0 e^{y/2+\ln \tau /\tau_0};|p|)\Bigr )} 
\;\;, \end{equation}
since $x\equiv\tau\sinh y/2$ and $t\equiv\tau\cosh y/2$ ($\tau\geq \tau_0>0$). 

Our results for free fermions show the free-streaming behavior of {\it classical dust}, 
associated with the independent propagation and linear superposition of the momentum  
contributions to the scalar component $F$ of the Wigner function here. In particular, 
the shape function of each mode is preserved and translated lightlike  
(with dispersion for massive particles). 
Due to the assumed momentum symmetry, the initial distribution will separate into two 
components after a finite time, travelling into the forward and 
backward direction, respectively, with a corresponding dilution at the center.  
    
We recall that $T^{\mu\nu}$ being diagonal implies 
the absence of ideal hydrodynamic flow, given $\epsilon =(d-1)P$. This does not depend 
on whether the initial state is on- or off-shell, see Eq.(\ref{TmunuEvolution}).   
Therefore, any hydrodynamic 
behavior must be the effect of a pecularity of the semiclassical limit \cite{DomLev00}, 
of coarse graining \cite{I95,Lisewski99,BrunHartle99},  
or of interactions \cite{Cooper01}, or a combination of these.  
   
Despite the apparently classical evolution, however, all initial state quantum effects are incorporated 
and preserved. If the 
initial dimensionless distribution $F$ has a dependence on products of 
momentum and space-time variables, which is characteristic for matter waves, such terms 
invoke a factor $1/\hbar$. Similarly, if it is thermal ($T$) but includes the 
finite size ($L$) shell effects or global constraints, then there are quantum 
corrections involving $LT/\hbar$ ($k_B=c=1$) \cite{IG}. They have not been 
included in semiclassical tranport or classical hydrodynamic models of high-energy 
(nuclear) collisions, but may be large. Here the quantum dust model provides  
a testing ground to assess the importance of these quantum effects. 
  
Our approach based on the Schwinger function reduces the solution of the 
free quantum transport problem to quadratures -- analytical results in three 
dimensions can be obtained and will be discussed elsewhere. It may lead 
to an efficient way of treating interacting particles, when a 
perturbative expansion is meaningful.  



\newpage
\begin{thebibliography}{99}

\bibitem{Fermi} E.\,Fermi, Progr.\,Theor.\,Phys. {\bf 5}, 570 (1950);
                          Phys.\,Rev. {\bf 81}, 683 (1951).

\bibitem{Landau} L.\,D.\,Landau, Izv.\,Akad.\,Nauk\,SSSR, Ser.\,fiz.,
                          {\bf 17}, 51 (1953);
                 S.\,Z.\,Belenkij and L.\,D.\,Landau, N.\,Cim., Suppl.,
                          {\bf 3}, 15 (1956).

\bibitem{Bj} J.\,D.\,Bjorken, Phys.\,Rev. {\bf D27}, 140 (1983).

\bibitem{QMs} See also the proceedings of the International Conferences on Ultra-Relativistic
Nucleus-Nucleus Collisions; e.g., {\it Quark Matter '99}, eds. L.\,Riccati, M.\,Masera and
E.\,Vercellini (Elsevier, Amsterdam, 1999).

\bibitem{Weinberg} S.\,Weinberg, {\it Gravitation and Cosmology}
                   (Wiley, New York, 1972).

\bibitem{DomLev00} G.\,\,Domenech and M.\,L.\,Levinas, Physica {\bf A278}, 440 (2000).

\bibitem{WHZ90} W.\,Zurek (ed.), {\it Complexity, Entropy and the Physics of Information}
                (Addison-Wesley, Reading, Mass., 1990).

\bibitem{I95} H.-Th.\,Elze, Nucl.\,Phys. {\bf B436}, 213 (1995);
              Phys.\,Lett. {\bf B369}, 295 (1996); [quant-ph/9710063].

\bibitem{Lisewski99} A.\,M.\,Lisewski, {\it On the classical hydrodynamic limit of quantum field theories}, 
                     [quant-ph/9905014].

\bibitem{BrunHartle99} T.\,A.\,Brun and J.\,B.\,Hartle, Phys.\,Rev. {\bf D60}, 123503 (1999).

\bibitem{EH} J.-P.\,Blaizot and E.\,Iancu, Nucl.\,Phys. {\bf B557}, 183 (1999); 
             D.\,Boyanovsky, F.\,Cooper, H.\,J.\,de\,Vega and P.\,Sodano, Phys.\,Rev. {\bf D58}, 025007 (1998);
             H.-Th.\,Elze and U.\,Heinz, Phys.\,Rep. {\bf 183}, 81 (1989).


\bibitem{VGE87} D.\,Vasak, M.\,Gyulassy and H.-Th.\,Elze, Ann.\,Phys. (N.Y.)
                {\bf 173}, 462 (1987).

\bibitem{neutrino} From this point on, a corresponding study of (approximately) massless Standard Model
                   $\nu_L\bar\nu_R$ neutrinos differs, and is simpler, since ${\cal V}_\mu ={\cal A}_\mu$\,,
                   while all other densities vanish identically; see, e.g., H.-Th.\,Elze, T.\,Kodama and
                   R.\,Opher, Phys.\,Rev. {\bf D63} (2001) 013008.

\bibitem{Madelung} E.\,Madelung, Z.\,Phys. {\bf 40} (1926) 322;
                   G.\,Holzwarth and D.\,Sch\"utte, Phys.\,Lett. {\bf B73} (1978) 255;
                   S.\,K.\,Ghosh and B.\,M.\,Deb, Phys.\,Rep. {\bf92} (1982) 1.

\bibitem{Cooper01} L.\,M.\,A.\,Bettencourt, F.\,Cooper and K.\,Pao, {\it Hydrodynamic scaling from 
the dynamics of relativistic quantum field theory}, [hep-ph/0109108].

\bibitem{IG} H.-Th.\,Elze and W.\,Greiner, Phys.\,Lett. {\bf B179}, 385 (1986); 
             Phys.\,Rev. {\bf A33},1879 (1986). 

\end{thebibliography}
\end{document}